\documentclass[twocolumn,showpacs,preprintnumbers,amsmath,amssymb]{revtex4}

\usepackage{graphicx}% Include figure files
\usepackage{dcolumn}% Align table columns on decimal point
\usepackage{bm}% bold math
\usepackage{epsfig}

\begin{document}

\preprint{Submitted to Phys. Rev. B }

\title{Tunable interactions between vortices and a magnetic dipole}
\author{Gilson Carneiro}
\affiliation{Instituto de F\'{\i}sica, Universidade Federal do Rio de Janeiro,  
C.P. 68528, 21941-972, Rio de Janeiro-RJ, Brasil }
 \email{gmc@if.ufrj.br}

\date{\today}

\begin{abstract}

The interactions between vortices in a thin superconducting film  and one magnetic dipole in the presence of a magnetic field applied parallel to the film surfaces are studied theoretically in the London limit. The dipole  magnetic moment is  assumed to have constant magnitude and freedom to rotate. The pinning potential for an arbitrary vortex configuration is calculated exactly. It is  found that, due to the dipole freedom to rotate, the pinning potential differs significantly from that for a permanent dipole. In particular,  its dependence on the applied field is non-trivial and allows for tuning of the pinning potential by the applied field. The critical current for one vortex pinned by the  dipole is obtained numerically as a function of the applied field and  found to depend strongly on the field. Order of magnitude changes in the critical current resulting from changes in the direction and magnitude of the applied field are reported, with discontinuous changes taking place in some cases. The effect of vortex pinning by random material defects on  the critical current is investigated using a simple model. It is found that if random pinning  is weak the critical current remains strongly dependent  on the applied field.  Possible applications to vortices pinned by arrays of magnetic dots are briefly considered.

\end{abstract}
\maketitle

%$$$$$$$$$$$$$$$$$$$$$$$$$$$$$$$$$$$$$$$$$$$$$$$$$$$$$$$$$$$$$$$$$$$$$$$$$$$4

\section{introduction}
\label{sec.int}

The study of interactions between vortices in superconducting films and arrays of magnetic dots  placed in the vicinity of the film has received a great deal of attention lately. The magnetic, superconducting, and transport  properties of a great variety of such systems  have been reported in the literature \cite{rev1,expp}. The main interest in this type of system is to enhance and modify vortex pinning, and thereby increase the critical current and stabilize new vortex phases. From the theoretical point of view, the  problem is to understand how the magnetization of the dots influence the vortices.  Calculations of the interactions between vortices and magnetic dots have been reported by several authors, using both the Ginzburg-Landau \cite{myp} and London theories \cite{coff,wei,sah,mypl,gmc1}. The experimental and theoretical work carried out in the above cited literature deals only with dots with permanent magnetization. In this case the interactions between vortices and magnetic dots result from the action of the inhomogeneous magnetic field created by the dots in the superconductor. As far as vortex pinning is concerned, the role played by the dots is a passive one, basically to set up a fixed pinning potential for the vortices. A question that naturally arises is what  happens if the dots have  non-permanent magnetization. In this case the dots are expected to play an active role in vortex pinning, in the sense that the magnetization of each dot now depends on the total magnetic field acting on it, which includes the magnetic field generated by the vortices and the applied field. The interaction between  vortices and dots must now differ from that for dots with permanent magnetization, and it may be even possible to tune it by changing the applied field.  

This paper studies theoretically the interaction between vortices and dots with non-permanent magnetization  using  a very simple model.  Vortices in a thin superconducting film interacting with one point dipole placed outside the film in the presence of an applied magnetic field parallel to the film surfaces. The dipole magnetic moment is assumed to have constant magnitude and freedom to rotate.  This model is also of experimental interest because  it is  feasible to fabricate magnetic dot arrays with freely rotating  magnetic moments, as demonstrated recently by Cowburn, Koltsov, Adeyeye, and Welland \cite{ckaw}. These authors reported on the magnetic properties of arrays of nanomagnets made of Supermalloy,  each nanomagnet being a thin circular disk of radius $R$. They found that for $R\sim 50-100$nm the magnetic state of each nanomagnet is a single domain one with the magnetization parallel to the disk plane, and that the magnetization can be reoriented by small applied fields. As pointed out in Ref.\ \onlinecite{ckaw}, each nonomagnet acts like a giant magnetic moment free to be oriented by magnetic fields acting on it. 

In this paper the pinning potential for an arbitrary  vortex configuration interacting with the dipole is calculated  exactly in the London limit, based on the  solutions of the London equations reported in 
Refs.\ \onlinecite{gmcehb,gmc1}.  It is found  that the pinning potential can be tuned by the applied field. The mechanism responsible for it is that the pinning potential depends on the dipole orientation which, in turn, depends on the applied field. The vortices are not influenced by the applied field because it is parallel to the film surfaces, and the film is thin. When a transport current is applied to the film, the magnetic field created by it  also contributes to the dipole orientation. This makes the pinning potential dependent on the transport current and, as shown here, has some important consequences for the critical current. It is also found that, in general, the pinning potential for many vortices is not simply the  sum  of the pinning potentials for each vortex. Applications for one an two vortices interacting with the dipole are considered. The pinning potentials are calculated, and their dependence on  the applied field is studied in detail. It is shown that the pinning potential can be tuned by the field over a wide range. 
The critical current for one vortex pinned by the dipole is then investigated.  It is found that the critical current depends strongly on the applied field. 
Changes in the critical current by as much as one order of magnitude are shown to result from changes in the magnitude and direction of the applied field. In the case of a magnetic moment parallel to the film surfaces, it is found that discontinuous jumps in the critical current take place in some circumstances, as the applied field magnitude is changed with the orientation fixed. The effect of random pinning on the critical current for the magnetic moment parallel to the film surfaces is investigated using a simple model. It is shown that if pinning is sufficiently weak  the  dependence of the critical current on the applied field is essentially unchanged, except that discontinuities are replaced by sharp changes. The paper also argues that the model under consideration is relevant for vortices interacting with arrays of nanomagnets  similar to those reported in Ref.\cite{ckaw}. 

This paper is organized as follows. In Sec.\ \ref{sec.vdi} the total energy for the superconductor-dipole system is obtained in the London limit and the pinning potential for an arbitrary vortex configuration is calculated. Applications to pinning of one and two vortices are also considered. In  
Sec.\ \ref{sec.jc} the critical current for one vortex pinned by the dipole is calculated numerically. Finally in Sec.\ \ref{sec.dsc} the results are discussed and the conclusions of the paper are stated. 

%############################################################################

\section{pinning potential}
\label{sec.vdi}

\begin{figure}[h]
\centerline{\includegraphics[scale=0.25]{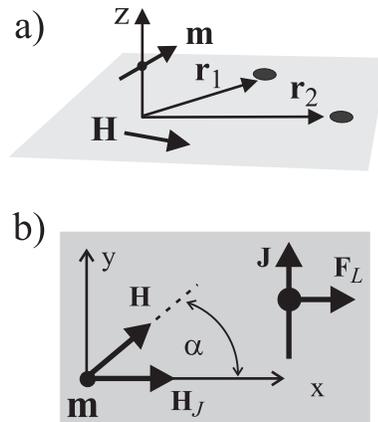}}
\vspace{5mm}
\caption {a) Superconducting film with two vortices  at  ${\bf r}_1$, and ${\bf r}_2$, a magnetic dipole, ${\bf m}$, at ${\bf r}_{0}=(0,0,z_0)$, and an applied magnetic field ${\bf H}$ parallel to the film 
surfaces. b) Film with a transport current ${\bf J}$ along the $y$-direction and one vortex. Transport current generates  field ${\bf H}_J$ at the dipole, and  force ${\bf F}_L$ on the vortex, both along the $x$-direction. ${\bf H}$ makes  an angle $\alpha$ with the $x$-axis.  }
\label{fig.fig1}
\end{figure}

The superconducting film is assumed to be planar, with surfaces parallel to each other and to the $x-y$ plane,  isotropic,  characterized by the penetration depth $\lambda$, and of thickness $d\ll \lambda$.  The film has a distribution of vortices, each vortex with vorticity $\,q_j=\pm 1\,$  and  located at positions ${\bf r}_j$.  The  magnetic dipole, ${\bf m}$, is located above the film at ${\bf r}_0 =(0,0,z_0>0)$, has constant magnitude $m$, but is free to rotate. 
Here two possibilities are considered: i) ${\bf m}$ parallel to the film surfaces ( $x-y$ plane), and  ii) ${\bf m}$ free to point in any direction in three-dimensional space. Hereafter these two possibilities are referred to as {\it parallel dipole} and {\it free dipole}, respectively. An uniform magnetic field ${\bf H}$ is applied parallel to the film surfaces.  The film-dipole system  is shown in Fig.\ \ref{fig.fig1}a.
%################################################################################# 
%######################################################################

In the London limit, the total energy can be written as
 \begin{equation}  
E_{T}= E_{sm}+E_v+E_{vm}+E_H \, ,
  \label{eq.ett}
  \end{equation} 
where $E_{sm}$ is the energy of interaction of the dipole with the magnetic  field created by the  
screening supercurrent induced by it in the film, $E_v$ is the total energy of the vortices in the absence of the dipole, $E_{vm}$ is the interaction energy of the dipole with the vortices, and $E_H=  - {\bf m}\cdot{\bf H}$ is the energy of the dipole in the applied field. It is convenient to use  the following natural scales for physical quantities. Length: $\xi=$ vortex core radius. Energy: $\epsilon_0d$,  where $\epsilon_0=(\phi_0/4\pi\lambda)^2$ is the basic scale for energy/length of the superconductor. Magnetic moment:  $\phi_0z_0$. Magnetic field: $\phi_0/\lambda^2$.

The  energy $E_{sm}$ is given by \cite{coff,wei,gmc1} 
 \begin{equation} 
 E_{sm}=-\frac{1}{2}{\bf m}\cdot {\bf b}'_{\bf m} \, ,
  \label{eq.esm}
  \end{equation}
where ${\bf b}'_{\bf m}$ is the field of the  screening supercurrent at the dipole position. For $d\ll \lambda$ 
  \begin{equation}  
{\bf b}'_{\bf m} = -\frac{d}{8\lambda^2z^2_0}(m_z{\bf \hat{z}} + \frac{{\bf m}_{\perp}}{2}) \, ,
  \label{eq.bpm}
  \end{equation} 
where $ {\bf m}_{\perp}$ is the component of $ {\bf m}$  perpendicular to the $z$-direction, that is, parallel to the film surfaces. 
For a parallel dipole, $E_{sm}$ is a constant, that is, independent of the dipole orientation, since 
$m_z=0,\; m_{\perp}=m$. For a free dipole, $E_{sm}$ depends only on the orientation of $ {\bf m}$  with respect to the $z$-axis. In this case the minimum of $E_{sm}$ occurs when the dipole is oriented parallel to the film surfaces ($m_z=0$), with  minimum energy $ E_{sm}= (\epsilon_0 d \pi^2/2) (m/\phi_0 z_0)^2$

The vortex energy $E_v$ is given by 
 \begin{equation}  
E_v= \epsilon_0d\,(\sum_i \, q^2_i \ln{\frac{\Lambda}{\xi}} + 
2\,\sum_{i\neq k} q_iq_k \ln{\frac{\Lambda}{\mid {\bf r}_i - {\bf r}_k \mid }}) \, ,
  \label{eq.evv}
  \end{equation}
where $\Lambda=2\lambda^2/d$, and  $\xi$ is the vortex core radius. The above expression is valid provided that $\mid {\bf r}_i - {\bf r}_k \mid  \ll \Lambda$ \cite{per}. 

The vortex-dipole energy, $E_{vm}$, results from the magnetostatic interaction of the dipole  with the magnetic field created by the vortices \cite{wei,sah,mypl,gmc1}, that is  
 \begin{equation}  
E_{vm}= - {\bf m}\cdot{\bf b} \, ,
  \label{eq.evm}
  \end{equation}  
where ${\bf b}$ is the field produced by the vortices at the dipole position. For single vortex with vorticity $q$ located at position  ${\bf r}$ such that $r \ll \Lambda$, the field, denoted by ${\bf b}^s$, is given by \cite{gmcehb}
 \begin{eqnarray} 
  {\bf b}^s_{\perp}  & = &-q\frac{\phi_0 d}{4\pi \lambda^2}\, \frac{{\bf r}}{r^2}(1-\frac{z_0}{\sqrt{r^2+z^2_0}})\, , \nonumber \\ 
  b^s_z & = &  q\frac{\phi_0 d}{4\pi \lambda^2}\frac{1}{\sqrt{r^2+z^2_0}} \, ,
  \label{eq.bvc} 
  \end{eqnarray}
where  ${\bf b^s_{\perp}}$ is the component of  ${\bf b}^s$ parallel to the film surfaces. For many vortices, ${\bf b}$ is given by the sum of the fields  ${\bf b}^s$  for each individual vortex. 

In summary then, the total energy can be written as 
 \begin{equation}  
E_{T}= E_v + E_{sm}- {\bf m}\cdot({\bf b} + {\bf H})\, ,
  \label{eq.ettb}
  \end{equation} 
and  depends both on the vortex positions and on the dipole orientation. The static interaction between the vortices and the dipole is the total energy for the dipole at its equilibrium orientation. That is, for  the direction of ${\bf m}$  that minimizes $E_{T}$ with the vortices held at fixed positions. For a parallel dipole, according to Eq.\ (\ref{eq.ettb}), the equilibrium ${\bf m}$ is parallel to ${\bf b}_{\perp}+{\bf H}$, because $E_{sm}$ is independent of the dipole orientation. For a free dipole, $E_{sm}$ also contributes to the equilibrium ${\bf m}$. In this paper the effect of $E_{sm}$ is neglected, so that the equilibrium ${\bf m}$ is parallel to ${\bf b}+{\bf H}$. The conditions for the validity of this approximation are discussed in more detail later. The total energy for the equilibrium  ${\bf m}$ can be written as  
\begin{equation}  
E^{eq}_{T}= E_v+U_{vm}\, ,
  \label{eq.ete}
  \end{equation} 
where $U_{vm}$ is the pinning potential for the vortices, given by
 \begin{equation}
U_{vm}=- m\mid {\bf b_{\perp}} +{\bf H}\mid+mH \; , 
  \label{eq.uvmp}
  \end{equation} 
 for a parallel dipole, and 
  \begin{equation}
U_{vm} =- m[ ({\bf b_{\perp}} +{\bf H})^2 + ( b_{z})^2)]^{1/2}+mH \; , 
  \label{eq.uvmf}
  \end{equation} 
for a free dipole. 
The constant term $mH$ is added to the definition of $U_{vm}$ in order that it vanishes when there are no vortices present, that is when ${\bf b}=0$.  Two important consequences of the dipole freedom to rotate present in Eqs.\ (\ref{eq.uvmp}) and (\ref{eq.uvmf}) are the many-vortex character of $U_{vm}$, 
and the non-trivial dependence of on ${\bf H}$. The former results because, according to Eqs.\ (\ref{eq.uvmp}) and (\ref{eq.uvmf}), $U_{vm}$ is not in general the sum of the pinning potentials for individual vortices. It depends on the vortex positions through ${\bf b}$, which is the sum of the individual vortex fields ${\bf b}^s$, Eq.\ (\ref{eq.bvc}). Thus, the dipole also generates vortex-vortex interactions. The dependence on ${\bf H}$  obtained in Eqs.\ (\ref{eq.uvmp}), and (\ref{eq.uvmf}) casts it in the role of a handle that controls the  strength and spatial dependence of $U_{vm}$. The pinning potentials for one and two vortices are discussed next.

\subsection{one vortex}
\label{sec.onv}

\begin{figure}[t]
\centerline{\includegraphics[scale=0.3]{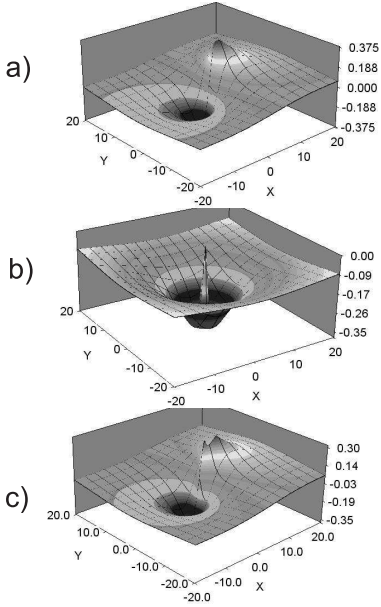}}
%\vspace{5mm}
\caption{Pinning potential for one vortex, $U^{1v}_{vm}$ (in units of $\epsilon_0 d$) interacting with  a parallel dipole. Parameters: $\lambda=10.0\xi, d=z_0= 2.0\xi$, and  $m=0.1\phi_0z_0$. External field in the $x$-direction: a) Permanent dipole , b) $H=0$ , c) $H=0.02\phi_0/\lambda^2$. X and Y in units of $\xi$. }
\label{fig.fig2}
\end{figure}

\begin{figure}[t]
\centerline{\includegraphics[scale=0.3]{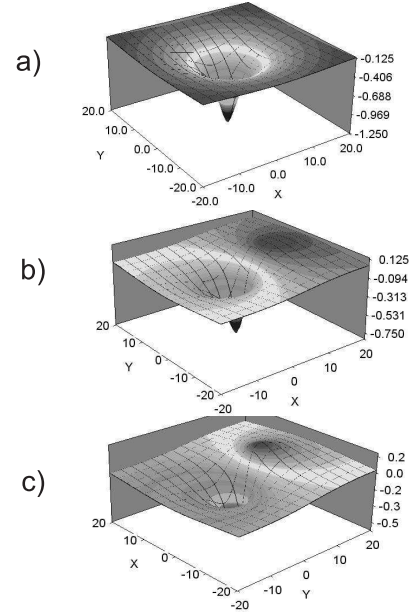}}
%\vspace{5mm}
\caption{Pinning potential for one vortex, $U^{1v}_{vm}$ (in units of $\epsilon_0 d$) interacting with a free dipole. Parameters: $\lambda=10.0\xi, d=z_0= 2.0\xi$, and $m=0.1\phi_0z_0$. External field in the $x$-direction: a) $H=0$, b)$H=0.04\phi_0/\lambda^2$, c)$H=0.1\phi_0/\lambda^2$ . X and Y in units of $\xi$ . }
\label{fig.fig3}
\end{figure}

The pinning potential for one vortex, denoted  by $U^{1v}_{vm}$, follows from Eqs.\ (\ref{eq.uvmp}), and (\ref{eq.uvmf}) with ${\bf b}$ replaced by ${\bf b}^s$, Eq.\ (\ref{eq.bvc}).
Typical results for the spatial dependence of $U^{1v}_{vm}$ for characteristic values of $H$, assuming that ${\bf H}$ is parallel to the $x$-axis, are  shown in Figs.\ \ref{fig.fig2} and  \ref{fig.fig3}.  The scale for $H$ in $U^{1v}_{vm}$ is  the vortex field, which is bound by 
$\mid b^s_{z}\mid \leq \mid q\mid\, d/4\pi z_0\times(\phi_0/\lambda^2) $, and 
$\mid {\bf b}^s_{\perp}\mid \leq 0.3\mid q\mid\, d/4\pi z_0\times(\phi_0/\lambda^2)$. The details of the dependence of $U^{1v}_{vm}$ on the spatial coordinates and on ${\bf H}$ are described next. 

\noindent i) $H\gg b^s$. In this case, since $H$ is much larger than the vortex field, the  equilibrium orientation of ${\bf m}$ is parallel to ${\bf H}$, and  $U^{1v}_{vm}$ reduces to the pinning potential for a  vortex interacting with a permanent dipole parallel to ${\bf H}$.    
Assuming that ${\bf H}$ is parallel to the $x$-direction, $U^{1v}_{vm}=-mb^s_{x}$, which, according to  Eqs.\ (\ref{eq.bvc}), coincides with   the expression obtained in Refs.\cite{wei,sah,mypl,gmc1}. In this case $U^{1v}_{vm}$ is anti-symmetric with  respect to an inversion of the vortex position (${\bf r} \rightarrow -{\bf r}$), and to the change of sign of the vorticity ($q \rightarrow -q$). For $q>0$ it  has a minimum ( maximum ) on the $x$-axis at  $x= - (+)1.3 z_0$, with  minimum (maximum) value  
$U^{1v}_{vm}/\epsilon_0 d=-(+)0.3\times 4\pi q(m/\phi_0 z_0)$, as shown in Fig.\ \ref{fig.fig2}.a.

\noindent ii){\it Parallel dipole.} For $H=0$, $U^{1v}_{vm}$ is given  by   
 \begin{equation} 
\frac{U^{1v}_{vm}}{\epsilon_0 d} = -4\pi \mid q\mid   \frac{m}{\phi_0 z_0}\, \frac{z_0}{r}
(1-\frac{z_0}{\sqrt{r^2+z^2_0 }}) \;. 
  \label{eq.eh0a}
 \end{equation}
Thus, for $H=0$, $U^{1v}_{vm}$  has circular symmetry, is the same for vortices ($q>0$) and anti-vortices ($q<0$), and is attractive with a repulsive core. The minimum of $U^{1v}_{vm}$ is  degenerate,  located on a circle of radius $r= 1.3 z_0$, centered at the dipole, and  minimum value  $U_{vm}/\epsilon_0 d=-0.3 \times4\pi\mid q\mid (m/\phi_0 z_0)$, as shown in Fig.\ \ref{fig.fig2}.b. For $H\neq0$, the spatial dependence of $U^{1v}_{vm}$ changes smoothly with $H$  
between  the $H=0$ and  $H\gg b^s$ limits, as shown in  Fig.\ \ref{fig.fig2}. The minimum of $U^{1v}_{vm}$ occurs when the vortex ($q>0$) is located on the negative $x$-axis. In this case ${\bf b}^s_{\perp}$ is parallel to ${\bf H}$, and  $U^{1v}_{vm} = - m b^s_x $, so that the minimum of $U^{1v}_{vm}$ is identical to that for a permanent dipole parallel to the positive $x$-direction. However, the  spatial dependence of $U^{1v}_{vm}$ differs  from that for a permanent dipole, as shown in Fig.\ \ref{fig.fig2}.c.

\noindent iii){\it Free dipole.} For $H=0$, $U^{1v}_{vm}$ is given  by
 \begin{equation}    
\frac{U^{1v}_{vm}}{\epsilon_0 d}  = -4\pi\mid q\mid \sqrt{2}   \frac{m}{\phi_0 z_0}\, \frac{z_0}{r}
(1-\frac{z_0}{\sqrt{z^2_0+ r^2}})^{1/2}\,.
  \label{eq.eh0b} 
  \end{equation}
The vortex pinning potential for $H=0$ also has circular symmetry, and is the same for vortices and anti-vortices. It is purely attractive with a minimum at $r=0$, and minimum value $U^{1v}_{vm}/\epsilon_0 d=-4\pi\mid q\mid (m/\phi_0 z_0)$, as shown in Fig.\ \ref{fig.fig3}.a. For $H\neq0$, the spatial dependence of $U^{1v}_{vm}$ changes with $H$ as shown in  Fig.\ \ref{fig.fig3}. The minimum of  
$U^{1v}_{vm}$ also occurs  when  ${\bf b}^s_{\perp}$ is parallel to ${\bf H}$. For $q>0$ it is  located on the negative $x$-axis, at a position that depends on $H$, between  $x=0$ ( $H=0$)  and $x=-1.3 z_0$ ( $H \gg b^s$). 
The minimum value of $U^{1v}_{vm}$ also depends on $H$, and decreases as $H$ increases from $U^{1v}_{vm}/\epsilon_0 d=-4\pi q (m/\phi_0 z_0)$ for $H=0$ to 
$U^{1v}_{vm}/\epsilon_0 d=-0.3\times 4\pi q(m/\phi_0 z_0)$ for  $H \gg b^s$. The approximation that neglects the contribution of $E_{sm}$ to the equilibrium orientation of the dipole in the case of a free dipole is justified  if $b^s >  b'_{\bf m}$. According  to Eqs.\ (\ref{eq.bpm}) and (\ref{eq.bvc}), this requires that 
$r/z_0\lesssim \mid q\mid \phi_0z_0/m$.
%
%################################################################################# 
%######################################################################
%
%################################################################################# 
%###################################################################### 
\subsection{two vortices}
\label{sec.twv}

The pinning potential  for two vortices, $U^{2v}_{vm}$, with vorticities $q_1$ and  $q_2$,  located, respectively, at ${\bf r}_1$ and ${\bf r}_2$, is given by
 \begin{equation}  
U^{2v}_{vm} = - m\mid {\bf b^s_{1\perp}}+ {\bf b^s_{2\perp}} +{\bf H}\mid  +mH \; , 
  \label{eq.evma}
\end{equation}
for the parallel dipole, and by 
\begin{equation}
U^{2v}_{vm} = - m\mid {\bf b}^s_{1}+ {\bf b}^s_{2} +{\bf H}\mid   
+mH \; ,
  \label{eq.evmb}
  \end{equation}
for the free dipole. In Eqs.\ (\ref{eq.evma}) and (\ref{eq.evmb})  ${\bf b}^s_1$ and ${\bf b}^s_2$ are, respectively, the fields of  vortices $q_1$ and $q_2$, given by Eqs.\ (\ref{eq.bvc}), with ${\bf r}$ replaced by  
${\bf r}_1$ and ${\bf r}_2$. In general $U^{2v}_{vm}$ is not the sum of the pinning potentials for each vortex. The exception occurs only  for $H$ much larger than the vortex fields. In this case  $U^{2v}_{vm}$ reduces to the energy of interaction of two vortices with a fixed dipole, equal to the sum of the  energies of interaction of each vortex with a dipole oriented along ${\bf H}$. The physical meaning  of  $U^{2v}_{vm}$  can be seen more clearly by calculating  the 
force exerted by the dipole on each vortex,  ${\bf F}_i=-{\mbox{\boldmath ${\nabla}$}}_iU^{2v}_{vm},\; i=1,2$. Using Eqs.\ (\ref{eq.evma}) and (\ref{eq.evmb}). The result  is 
 \begin{equation}  
 F_{ix}={\bf m}_{eq}\cdot\frac{\partial {\bf b}^s_i}{\partial x_i},\;\; 
 F_{iy}={\bf m}_{eq}\cdot\frac{\partial {\bf b}^s_i}{\partial y_i} \; .
   \label{eq.fix}
\end{equation}
In Eq.\ (\ref{eq.fix}) ${\bf m}_{eq}$ denotes the equilibrium magnetic moment, that is, $\mid {\bf m}_{eq}\mid =m$ and orientation parallel to the total field acting on the dipole, namely  
\begin{equation}
{\bf m}_{eq}=m\frac{{\bf b}^s_{1\perp}+ {\bf b}^s_{2\perp} +{\bf H}}{\mid {\bf b}^s_1+ {\bf b}^s_2 +{\bf H}\mid}\; , 
  \label{eq.meqp}
  \end{equation} 
 for the parallel dipole, and  
\begin{equation}
{\bf m}_{eq}=m\frac{{\bf b}^s_1+ {\bf b}^s_2 +{\bf H}}{\mid {\bf b}^s_1+ {\bf b}^s_2 +{\bf H}\mid}\; , 
  \label{eq.meqf}
  \end{equation}   
  for the free dipole. 
For a permanent dipole, the expression for ${\bf F}_i$ is identical to  Eq.\ (\ref{eq.fix})  with ${\bf m}_{eq}$ replaced by the permanent dipole moment. Thus, the force on each vortex is the same as that exerted by a permanent dipole with magnetic moment ${\bf m}_{eq}$. However, the force on one vortex also depends on the position of the other vortex, through ${\bf m}_{eq}$. This shows that the dipole freedom to rotate generates vortex-vortex interactions.  

As an application of the above results, pinning of two vortices with $q_1=q_2=1$ by the dipole is now considered. The equilibrium positions of the two vortices  minimize the total energy, given by  Eq.\ (\ref{eq.ettb})  with $U_{vm}$ replaced by $U^{2v}_{vm}$, Eqs.\ (\ref{eq.evma}) and (\ref{eq.evmb}).  The energy  $E_{sm}$ is neglected in what follows. 
%################################################################################# 
\begin{figure}[t]
\centerline{\includegraphics[scale=0.3]{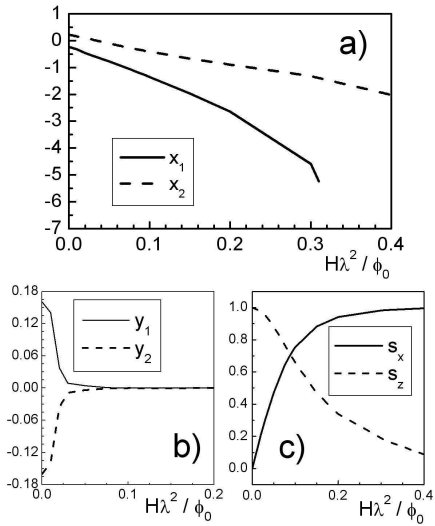}}
\caption{Panels a) and b): positions (in units of $\xi$) of a vortex pair $q_1=q_2=1$ pinned by the a free dipole. In a) only one vortex is pinned for $H>0.31\phi_0/\lambda^2$.   c) magnetic moment orientation (${\bf s}={\bf m}/m\;, s_y=0$). Parameter values:  $\lambda=10.0\xi, d=z_0= 2.0\xi$ , and $m=\phi_0z_0$.  }
\label{fig.fig4}
\end{figure}
%######################################################################

\noindent {\it Parallel dipole.} The total energy it is minimized when  the vortex fields ${\bf b^s_{1\perp}}$ and  ${\bf b^s_{2\perp}}$ are parallel to each other and to ${\bf H}$. For ${\bf H}$ along the $x$-direction, both vortices are  located on the $x$ axis. In this case $U^{2v}_{vm}$, Eq.\ (\ref{eq.evma}), becomes 
$U^{2v}_{vm}=- m( b^s_{1x}+  b^s_{2x})$, which is identical to the energy of interaction of two vortices located on the $x$-axis with a permanent dipole parallel to the $x$-axis. Thus the equilibrium positions of the vortices are independent of $H$, and identical to those for a permanent dipole.  For $H=0$ the minimum of the total energy  also occurs when the two vortices  are collinear with the dipole position in the $x-y$ plane, but their positions are degenerate with respect to the orientation of the line joining the vortices.

\noindent {\it Free dipole.} It follows from Eq.\ (\ref{eq.evmb}) that the equilibrium positions for $H=0$ are related by inversion symmetry with respect to the  dipole, that is   $x_2=-x_1,\; y_2=-y_1$. This arrangement cancels the component of the vortex field parallel to the film surfaces, so that the dipole is oriented along the $z$-direction. When $H\neq 0$ this symmetry is broken, and the vortex positions depend on $H$. Numerical results  are shown Fig.\ \ref{fig.fig4}. As $H$ increases  the two vortices approach the $x$-axis, and eventually become located on it ($H>0.08\phi_0 /\lambda^2$ in Fig.\ \ref{fig.fig4}.b). For the parameters used in Fig.\ \ref{fig.fig4}, only one vortex is pinned by the dipole for large $H$ ($H>0.31\phi_0 /\lambda^2$ in Fig.\ \ref{fig.fig4}.a). The other vortex depins  at $H=0.31\phi_0 /\lambda^2$, and is repelled away from the dipole. The orientation of the  dipole changes smoothly with $H$,  being nearly parallel to the film surfaces when one of the vortices depins (Fig.\ \ref{fig.fig4}.c).

\section{critical current}
\label{sec.jc}
Now  the critical current, $J_c$, for one vortex pinned by the dipole is considered. The effect of a  transport current density, ${\bf J}$,  applied to the film is twofold. First, it exerts on the vortex a force  ${\bf F}_L=(\phi_0 d/c){\bf J}\times \hat{{\bf z}}$. Second, it  creates a field at the dipole position ${\bf H}_J=(2\pi d/c){\bf J}\times \hat{{\bf z}}$, which adds to the external field ${\bf H}$, and modifies the vortex pinning potential. The reason is that $U^{1v}_{vm}$ is now given by Eqs.\ (\ref{eq.uvmp}) and  (\ref{eq.uvmf}) with ${\bf H}$ replaced by the total field ${\bf H}_{T}={\bf H}+{\bf H}_J$. The critical current depends on the relative orientation of  ${\bf J}$ and ${\bf H}$. Here it is assumed that ${\bf J}$ is fixed in the positive $y$-direction, and that ${\bf H}$ points in  a direction that makes an angle $\alpha $  with the positive $x$-axis ( see Fig.\ \ref{fig.fig1}.b). In this case both ${\bf F}_L$ and ${\bf H}_J$ are along the positive $x$-direction, and have magnitudes $H_J= 2\pi dJ /c$ and  $F_L=\phi_0dJ/c$. In this paper $J_c$ is obtained by solving numerically  the equations of motion for the vortex. It is assumed that for ${\bf J}=0$ the vortex is pinned at the absolute minimum of $U^{1v}_{vm}$, and that $J$ increases very slowly with time. These assumptions ensure that the vortex follows the position of  the minimum of $U^{1v}_{vm}-  F_L\,x$ as $J$ increases, until $J$ reaches a value for which the  minimum becomes unstable, and the vortex depins. As $J$ increases further, the vortex velocity also increases. The value of  $J_c$ obtained here corresponds to $J$ for which the vortex velocity reaches a small value chosen for numerical convenience. The obtained $J_c$ is slightly larger than  $J$ for which the minimum becomes unstable. This is analogous to the voltage criterion in  $J_c$ measurements.  The values of $J$ are, of course, limited to $J<J_d$, where $J_d=c\phi_0/(12\sqrt{3}\pi^2\lambda^2\xi)$ is the depairing current. In the results reported next, regions where $J_c>J_d$ are  discussed for the sake of completeness. Now that $H_J$ also enters in 
$U^{1v}_{vm}$, it must be compared with the vortex field. The maximum $H_J$ occurs for $J=J_c$, and can be written as $H_{J_c}=0.031d/\xi(J_c/J_d)(\phi_0/\lambda^2)$. For $J_c\sim J_d$ and $d\sim z_0\sim\xi$, $H_{J_{c}}$ is comparable to the maximum value of $b^s_{\perp}$ and less than the maximum value of $ b^s_z$. As discussed next,  ${\bf H}_J$ affects the parallel dipole more than the free dipole. 

\begin{figure}[t]
\centerline{\includegraphics[scale=0.35]{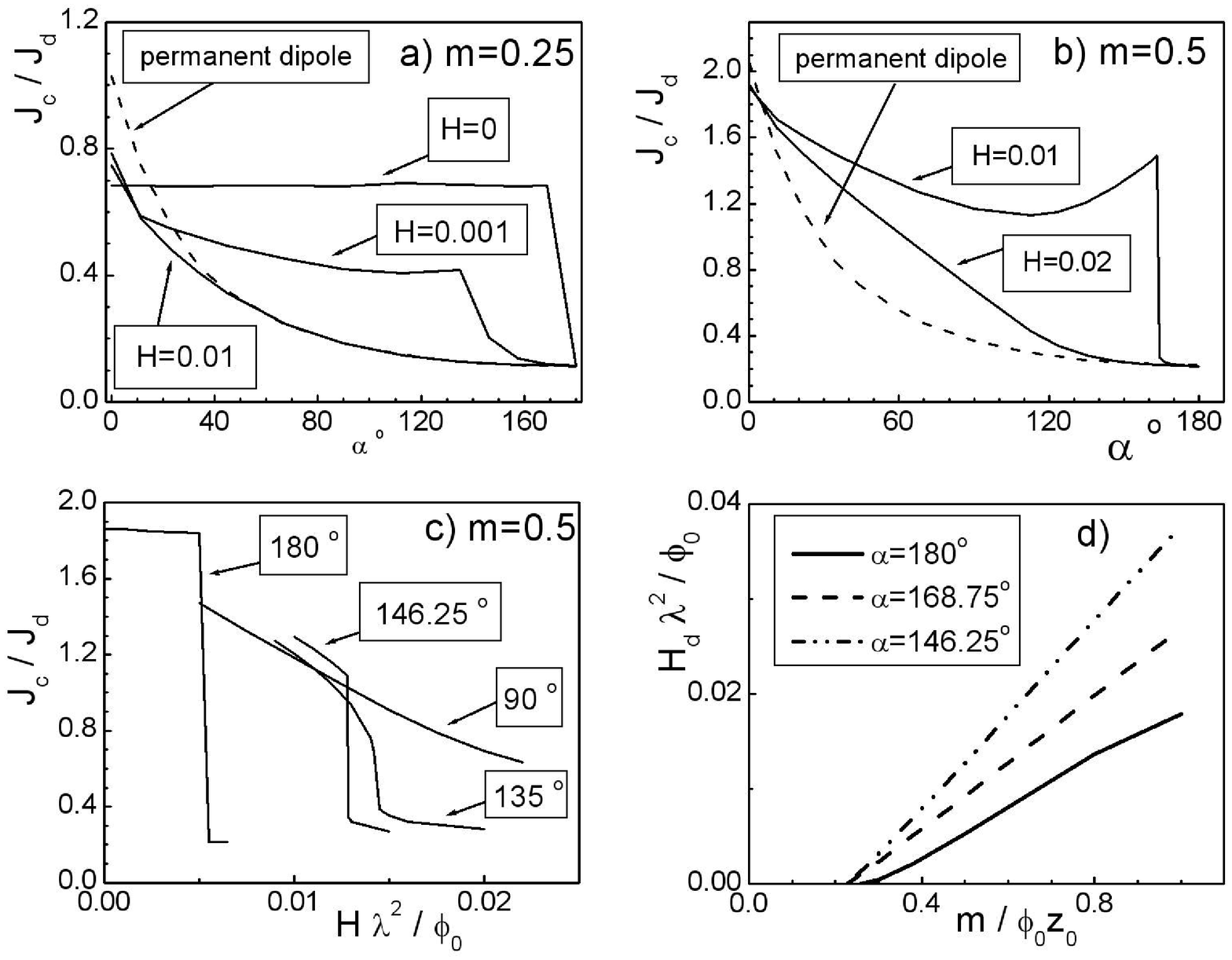}}
\vspace{5mm}
\caption{Single vortex  pinned by a parallel dipole for  $\lambda=10\xi, d=z_0= 2\xi$: a) and b) Critical current ($J_c$) vs. $\alpha$;  c) $J_c$ vs. $H$; d) discontinuity field $H_d$ vs $m$. Labels: $m$ in units of $\phi_0z_0$,  $H$ in units of  $\phi_0/\lambda^2$. }
\label{fig.fig5}
\end{figure}

\begin{figure}[t]
\centerline{\includegraphics[scale=0.2]{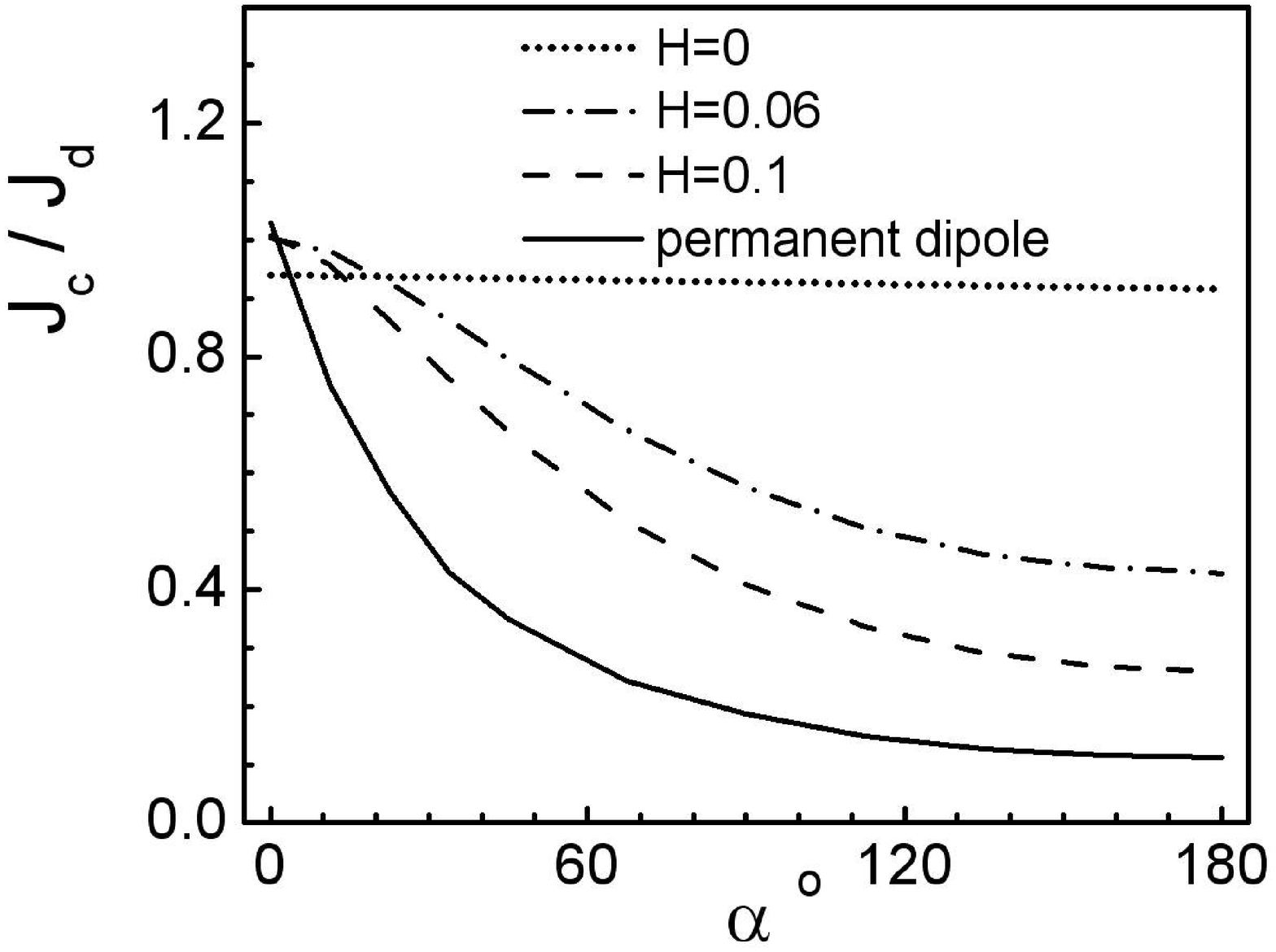}}
%\vspace{5mm}
\caption{Critical current ($J_c$) vs. $\alpha$ for one vortex pinned by a free dipole. Parameters: $m=0.5\phi_0z_0$ and $\lambda=10.0\xi, d=z_0= 2.0\xi$. Labels:  $H$ in units of $\phi_0/\lambda^2$. }
\label{fig.fig6}
\end{figure}

Typical  results for the dependence of $J_c$ on  $\alpha$ and $H$ are shown in  Fig.\ \ref{fig.fig5} and 
Fig.\ \ref{fig.fig6}. For  $H$ much larger than the vortex field and to  $H_J$,   the pinning potential  reduces to that for a permanent dipole oriented parallel to ${\bf H}$, that is,  at an angle $\alpha $ with the $x$-axis. The corresponding critical currents are shown in Fig.\ \ref{fig.fig5} and Fig.\ \ref{fig.fig6} by the curves labeled permanent dipole. For $H$ comparable to the vortex field and to $H_J$, the critical currents 
differ considerably from those for a permanent dipole. Most remarkable is  $J_c$  for the parallel dipole.  The $J_c$ vs. $\alpha$ curves (Figs.\ \ref{fig.fig5}.a and \ref{fig.fig5}.b) shown sharp jumps close to $\alpha=180^o$. The  $J_c$ vs. $H$ curves has sharp jumps and discontinuities (Fig.\ \ref{fig.fig5}.c). The reasons for the above described behavior are explained in detail next.
%
%################################################################################# 
%######################################################################
%
%################################################################################# 
%######################################################################

\noindent {\it Permanent dipole} ( $H\gg b^s,H_J$). The vortex pinning potential for a permanent dipole oriented at an angle $\alpha $ with the $x$-axis has spatial dependence like that shown in Fig.\ \ref{fig.fig2}.a, rotated by $\alpha$ with respect to the $x$-axis, and is independent of $H$ and $J$. For $J=0$, the vortex is pinned at the absolute minimum of $U^{1v}_{vm}$, located at a point in the $x-y$ plane defined in polar coordinates, $(\rho,\theta)$, by $(\rho=1.3z_0,\; \theta=\alpha +\pi)$. The critical current depends on $\alpha$ and $m$, being a linear function of $m$, since $U^{1v}_{vm}$ is linear in $m$. It is found that $J_c$ depends strongly on $\alpha$, being largest for  $\alpha=0$, and decreasing  smoothly with $\alpha$, as shown in Fig.\ \ref{fig.fig5} and  \ref{fig.fig6}. This results from the spatial dependence of $U^{1v}_{vm}$, as can be seen for $\alpha=0,\;180^o$. In these cases the critical current can be calculated analytically,  because the vortex moves only along the $x$ direction as $J$ increases. The result is  $J_c/J_d\simeq 4m/\phi_0z_0$   for $\alpha=0$, and  
$J_c/J_d\simeq 0.4m/\phi_0z_0$ for  $\alpha=180^o$.  The origin of this tenfold  difference can be seen in the plot of $U^{1v}_{vm}$ shown  Fig.\ \ref{fig.fig2}.a.  The driving force  is parallel  to the  $x$-axis in Fig.\ \ref{fig.fig2}.a for $\alpha=0$,  and antiparallel for $\alpha=180^o$. As can be seen in  Fig.\ \ref{fig.fig2}.a, the slope of the potential barrier is much steeper in the positive $x$-direction than in the negative one. For other values of $\alpha$, the depinning process is more complicated because the vortex motion as $J$ increases is not  confined to the direction of drive.

\noindent{\it Parallel dipole:} Typical results are shown in Fig.\ \ref{fig.fig5}. The $J_c$ vs. $\alpha$ curves are shown in Fig.\ \ref{fig.fig5}.a  for $m=0.25\phi_0z_0$, and in  Fig.\ \ref{fig.fig5}.b for $m=0.5\phi_0z_0$,  for characteristic  values of $H$. In both cases the $J_c$ vs. $\alpha$ curves differ considerably from those for a fixed dipole, being strongly dependent on $H$. There are sharp changes in $J_c$ close to $\alpha=180^o$, like those  for  $m=0.25\phi_0z_0$, $H=0.001\phi_0/\lambda^2$ (Fig.\ \ref{fig.fig5}.a) and  $m=0.5\phi_0z_0$, $H=0.01\phi_0/\lambda^2$ (Fig.\ \ref{fig.fig5}.b). The curve labeled $H=0$ in Fig.\ \ref{fig.fig5}.a is the limit of $J_c$ vs. $\alpha$ curve as $H\rightarrow 0$ with $\alpha$ fixed. The strong dependence of $J_c$ on $H$  is even more evident if $J_c$ is plotted as a function of  $H$ for  fixed $\alpha$,  as shown in Fig.\ \ref{fig.fig5}.c for $m=0.5\phi_0z_0$. In this case it is found that for  $\alpha \geq 146.25^o$ the $J_c$ vs. $H$ curves have discontinuities at $H=H_d$.  For  $\alpha < 146.25^o$, the dependence of $J_c$ on $H$ is continuous, as illustrated by the curves  for $\alpha=135^o$ and $\alpha=90^o$. For $\alpha=135^o$,  $J_c$ undergoes a rapid change with $H$ around  $H=0.014\phi_0/\lambda^2$, whereas for $\alpha=90^o$ the change in $J_c$ with $H$ is much slower.   It is found that the $J_c$ vs. $H$ curves have no discontinuities if $m$ is smaller than  a minimum value  which depends on $\alpha$. As shown in Fig.\ \ref{fig.fig5}.d,  $H_d$ vanishes at the minimum $m$,  and increases above it  essentially linearly with $m$. The reasons for  the sharp changes in the $J_c$ vs. $\alpha$ curves, and for the discontinuities in the $J_c$ vs. $H$ curves are explained in detail in what follows. 

\begin{figure}[t]
\centerline{\includegraphics[scale=0.35]{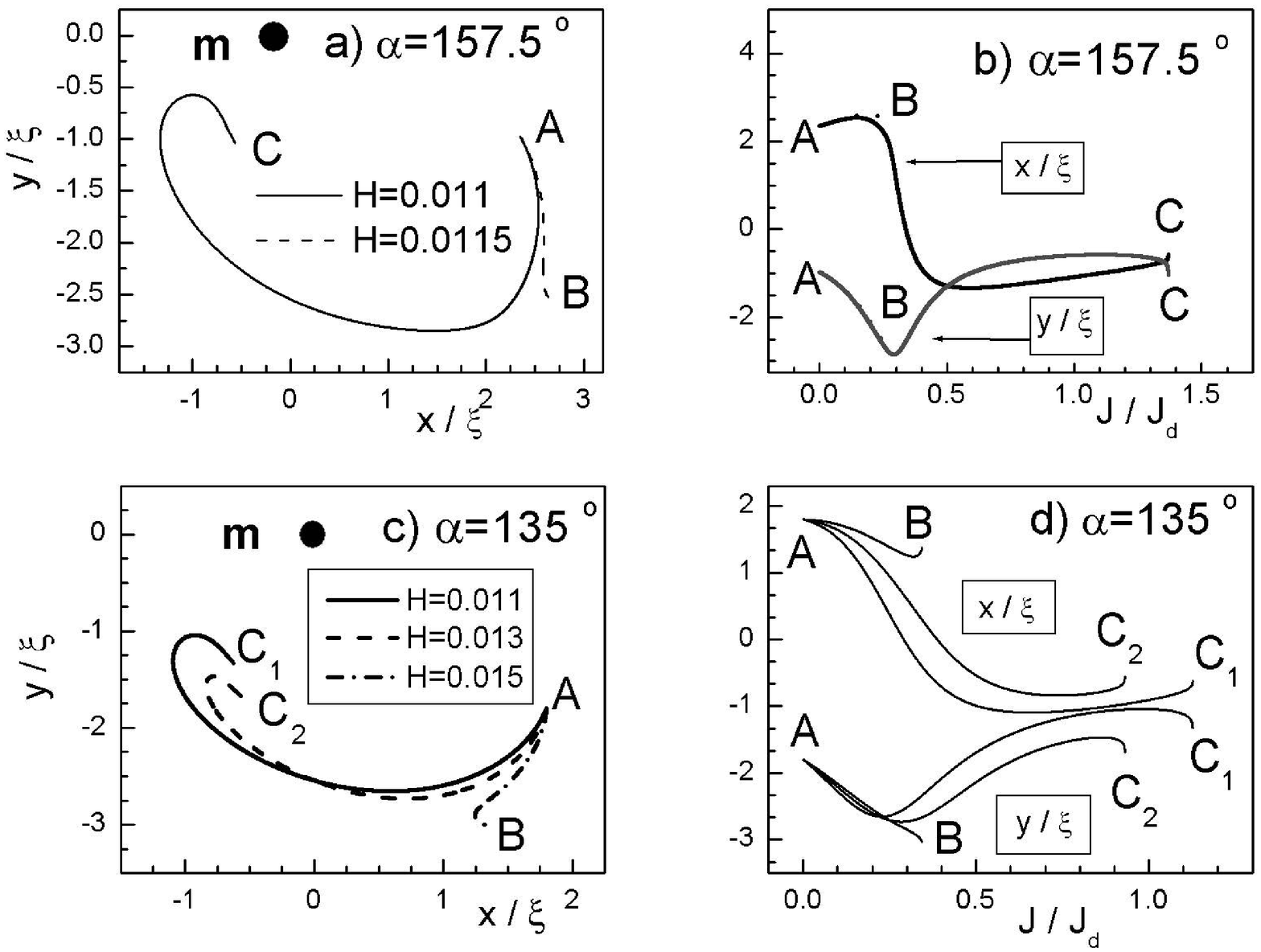}}
%\vspace{5mm}
\caption{Vortex pinned by a parallel dipole for $m=0.5\phi_0z_0$, $\lambda=10.0\xi$, and $ d=z_0= 2.0\xi$. a) and c): Vortex trajectories. Dot indicates dipole location in $x-y$ plane. b) and d): Vortex coordinates $x,\,y$ vs. $J$ corresponding to trajectories shown in a) and c). In d) top curves represent $x/\xi$, bottom curves $y/\xi$. Labels:  A = initial vortex position ( $J=0$). B and C = positions where the vortex depins.  $H$ in units of $\phi_0/\lambda^2$}
\label{fig.fig7}
\end{figure}

\noindent{\it $J_c$ vs. $\alpha$ curves}. The sharp jumps close to $\alpha=180^o$ result from the dependence of $U^{1v}_{vm}$ on $J$, through ${\bf H}_{T}={\bf H}+{\bf H}_J$. Basically what happens is that as $J$ increases, ${\bf H}_T$ rotates, becoming   nearly parallel to positive $x$-direction when $J$ reaches $J_c$. The vortex pinning potential changes accordingly. When $J$ reaches $J_c$ it is essentially identical to $U^{1v}_{vm}$ for a permanent dipole oriented close to the positive $x$-direction,  giving rise to a large $J_c$. To  demonstrated this, the instantaneous vortex position, which coincides with the minimum of $U^{1v}_{vm}- F_L x$, is studied as a function of $J$ as it increases with time for  $m=0.5\phi_0z_0$ and for typical values of $\alpha$. The values of $\alpha$ chosen are  $\alpha=157.5^o,\; 135^o$, for which the $J_c$ vs. $H$ curves have, respectively, a discontinuity at $H_d=0.0113\phi_0/\lambda^2$, and  no discontinuity. The trajectories in the $x-y$ plane described by the vortex are shown in Figs.\ \ref{fig.fig7}.a and  \ref{fig.fig7}.c. The vortex $x$ and $y$ coordinates are shown as a function of $J$ in  
Figs.\ \ref{fig.fig7}.b and  \ref{fig.fig7}.d. For $\alpha=157.5^o$ and $H=0.011\phi_0/\lambda^2<H_d$, and for  $\alpha=135^o$ and $H=0.011\phi_0/\lambda^2,\, 0.013\phi_0/\lambda^2$,  $J_c$ is enhanced by almost one order of magnitude with respect to the permanent dipole one. In these cases the position of the minimum  undergoes a large displacement, from the initial one on the right side of the dipole (A in Fig.\ \ref{fig.fig7}) for $J=0$, to the final one, on the left side of the dipole (C in Fig.\ \ref{fig.fig7}) for $J=J_c$. This is accompanied by a flip in the direction of ${\bf H}_{T}$ from near the negative $x$-direction  at $J=0$ to one near the positive $x$ direction for $J=J_c$. For  $\alpha=157.5^o$ and $H=0.011\phi_0/\lambda^2$, for instance,  $J_c=1.35J_d$, and  ${\bf H}_{T}$ points at $3^o$ with the $x$-axis and has magnitude $H_{T}=0.074\phi_0/\lambda^2$. The enhancement in $J_c$ results because the vortex is effectively pinned by a permanent dipole oriented at a small angle with the positive $x$-axis, as can be seen from the results for  $\alpha=157.5^o$ and $H=0.011\phi_0/\lambda^2$,  shown in Fig.\ \ref{fig.fig7}.b. Most of the displacement of the minimum takes place for $0<J/J_d<0.5$. For $J/J_d=0.5$, ${\bf H}_{T}$ points at $12^o$ with the positive $x$-axis and has magnitude $H_{T}=0.021\phi_0/\lambda^2$. The corresponding  $U^{1v}_{vm}$is essentially identical to that for a permanent dipole oriented at $12^o$ with the $x$-axis. For $0.5<J/J_d<1.35$, the change in  ${\bf H}_{T}$  has only a small effect on  $U^{1v}_{vm}$. Thus the vortex is effectively pinned by a permanent dipole oriented at an angle between  $12^o$ and $3^o$ with the $x$-axis. The resulting $J_c$ is comparable with that for a permanent dipole with this orientation.     
For  $\alpha=135^o$ and $H=0.011\phi_0/\lambda^2,\, 0.013\phi_0/\lambda^2$, a similar mechanism leads to the $J_c$ enhancement, as shown in Fig.\ \ref{fig.fig7}.c and \ref{fig.fig7}.d. For $\alpha=157.5^o$ and $H=0.0115\phi_0/\lambda^2>H_d$, and for  $\alpha=135^o$ and $H=0.015\phi_0/\lambda^2$,  on the other hand, where there is no enhancement in  $J_c$, the position of the minimum undergoes only a small displacement, from A to B in  Fig.\ \ref{fig.fig7}. For $\alpha=157.5^o$ and $H=0.0115\phi_0/\lambda^2$, it is found that the minimum becomes unstable at $J=0.25J_d$ and that  the corresponding ${\bf H}_{T}$  has magnitude $H_{T}=0.007\phi_0/\lambda^2$ and makes an angle of $50^o$ with the positive $x$-axis. 
%
%################################################################################# 
%######################################################################
%

\begin{figure}[t]
\centerline{\includegraphics[scale=0.3]{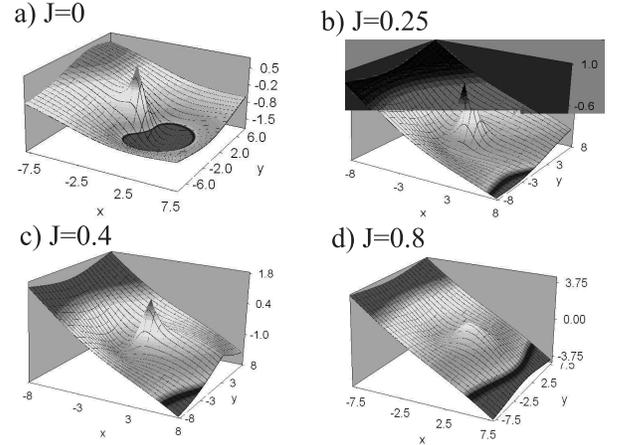}}
%\vspace{5mm}
\caption{Plot of $U^{1v}_{vm}- F_L x$ for parallel dipole  with increasing $J$. Parameters:  $m=0.5\phi_0z_0$ and $\lambda=10.0\xi, d=z_0= 2.0\xi$, $H=0.0113\phi_0/\lambda^2$, $\alpha=157.5^o$.  Labels:  $J$ in units of $J_d$. }
\label{fig.fig8}
\end{figure}

\noindent{\it $J_c$ vs. $H$ curves}. The reason for the discontinuities in the   $J_c$ vs. $H$ curves can be seen from the results for $\alpha=157.5^o$. The vortex equilibrium positions for $H=0.0115\phi_0/\lambda^2>H_d$ and for $H=0.011\phi_0/\lambda^2<H_d$, shown in Figs.\ \ref{fig.fig7}.a and \ref{fig.fig7}.b,  essentially coincide from A to B. At B the vortex depins if  $H=0.0115\phi_0/\lambda^2>H_d$, but if $H=0.011\phi_0/\lambda^2<H_d$ the vortex only depins at C. However, since $U^{1v}_{vm}- F_L x$ is a continuous function, the positions of the minima for two $H$ values so close to each other cannot differ much from one another. What happens is that the minimum for $H=0.011\phi_0/\lambda^2<H_d$ becomes unstable at B, when  $J=0.25J_d$, and a stable minimum appears again, at a slightly larger value of $J$. This new minimum  follows closely the position of the $H=0.011\phi_0/\lambda^2$ minimum,  becoming unstable close to point C.  However, the vortex depins when the minimum becomes unstable for the first time at point B. The evolution of $U^{1v}_{vm}- F_L x$ with $J$ is shown in Fig.\ \ref{fig.fig8}. The $J=0$ minimum  (Fig.\ \ref{fig.fig8}.a) becomes unstable at  $J=0.25J_d$ (Fig.\ \ref{fig.fig8}.b), but for $J=0.4J_d,\, 0.8J_d$  stable minima are present again 
(Fig.\ \ref{fig.fig8}.c and \ref{fig.fig8}.d). For $H=0.0115\phi_0/\lambda^2>H_d$, the minimum only becomes unstable once at point C.   Thus, the discontinuities in $J_c$ result because for $H>H_d$ the minimum becomes unstable twice, whereas for $H<H_d$ it becomes unstable only once. For $\alpha=135^o$ there are no discontinuities in $J_c$ vs. $H$, because the minimum only becomes unstable once for all $H$. The same is true for smaller values of $\alpha$. 
%################################################################################# 
%######################################################################

\noindent{\it Free dipole:} In this case the $J_c$ vs. $\alpha$ curves are smooth. The same is true for the  $J_c$ vs. $H$ curves. The results for  $m=0.25\phi_0z_0$, $\lambda=10.0\xi$, and $ d=z_0= 2.0\xi$ are shown in Fig.\ \ref{fig.fig6}. Large changes in $J_c$ with $H$ still take place,  except for small $\alpha$.  For $H=0$, $J_c$ is, of course,  independent of $\alpha$. For  $J_c\sim J_d$,  $H_T\sim 0.03\phi_0/\lambda^2$, and the maximum vortex field is $0.08\phi_0/\lambda^2>H_T$. As $H$ increases, $J_c$ approaches the permanent dipole limit. Note, however, that the value of $H$ needed to reach this limit is larger than that for a parallel dipole, due to the effect of the $z$-component of the vortex field, as discussed in Sec.\ \ref{sec.vdi}.

\subsection{random pinning}
\label{sec.rdp}
Vortex pinning by random material defects can modify the strong dependence of  $J_c$ on $H$ obtained above for the parallel dipole. The reason is that in the presence of random pinning the vortex equilibrium position no longer coincides with  the minimum of  $U^{1v}_{vm}- F_L x$. The vortex motion as $J$ increases is thus changed, modifying the dependence of  $J_c$ on $H$. It is expected that if the random pinning force is small compared with the dipole pinning force, $J_c$ remains essentially identical to that in the absence of pinning. Here the question of how large the random pinning force must be in order that the strong dependence of $J_c$ on $H$  is destroyed is investigated.

\begin{figure}[h]
\centerline{\includegraphics[scale=0.3]{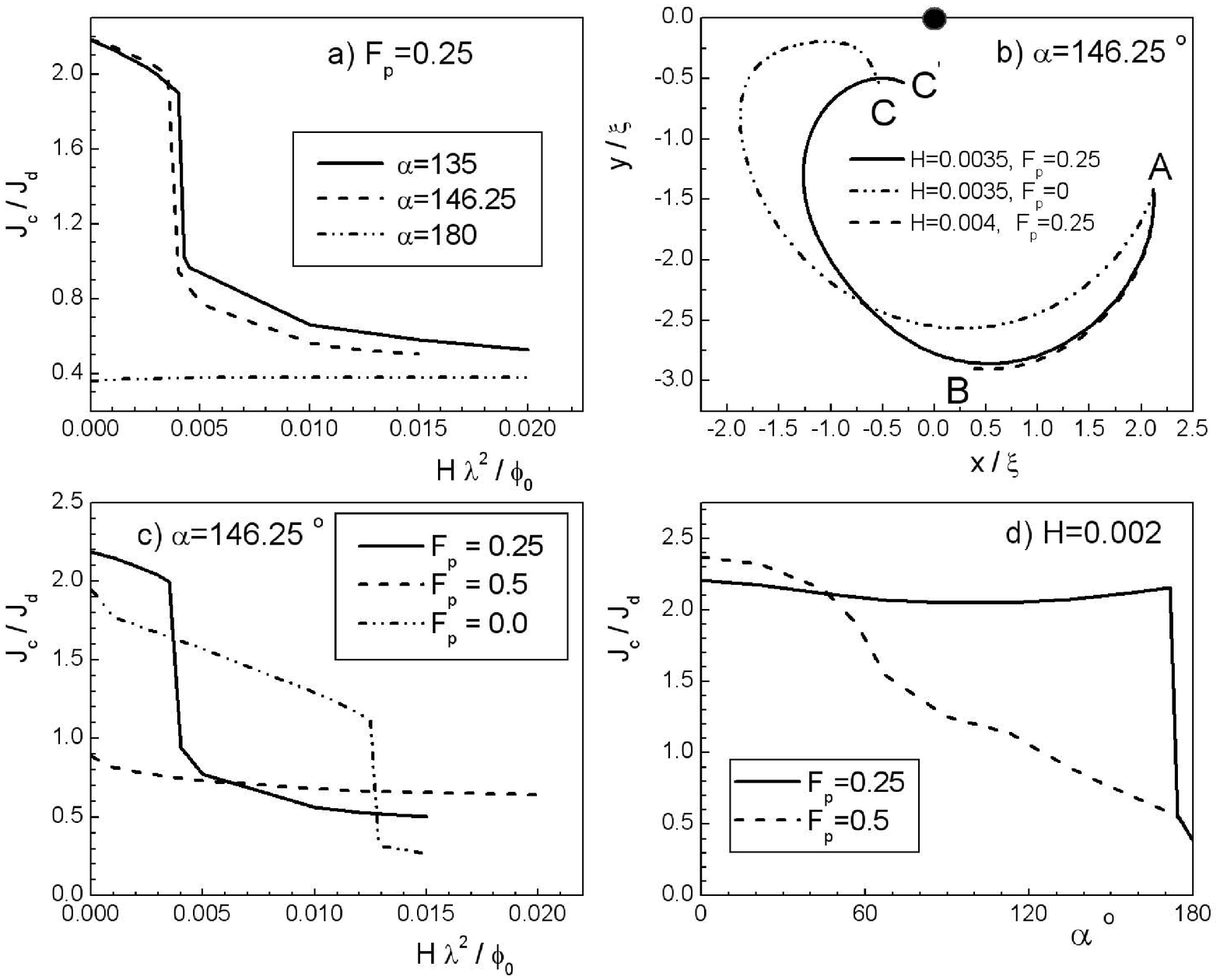}}
%\vspace{5mm}
\caption{Vortex pinned by a parallel dipole in the presence of random pinning. a) $J_c$ vs. $H$ curves for several 
$\alpha$.  b) Vortex trajectories with and without random pinning. Labels: A= vortex initial location ($J=0$). B and C= locations where vortex depins for $H$ and $F_p$ as indicated. c) and d) $J_c$ vs. $H$ and  $J_c$ vs. $\alpha$ curves for various pinning strengths. Parameters: $m=0.5\phi_0z_0$, $\lambda=10.0\xi$, and $d=z_0= 2.0\xi$:   $H$ in units of $\phi_0/\lambda^2$, $F_p$ in units of $\epsilon_0$. }
\label{fig.fig9}
\end{figure}

Random pinning is incorporated in the calculation of $J_c$ by the following  simple model. The random pinning force magnitude, $F_p$, is assumed  given, and independent of $H$ and $J$. If the total force exerted on the vortex by the dipole and by the transport current, ${\bf F}=-{\mbox{\boldmath ${\nabla}$}}U^{1v}_{vm}+{\bf F}_{L}$,  is such that  $F<F_p$, the vortex remains pinned. If, on the other hand, $F>F_p$ the vortex moves, and the total force acting on it is ${\bf F}(1-F_p/F)$. The vortex equations of motion are now  solved including the random pinning force, and  with the same initial conditions as above. That is, with the vortex pinned at the absolute minimum of  $U^{1v}_{vm}$ for $J=0$. The $J_c$ vs. $H$ curves  for $m=0.5\phi_0z_0$,  $\lambda=10.0\xi, d=z_0= 2.0\xi$, and $F_p=0.25\epsilon_0,\; 0.5\epsilon_0$  are shown in Fig.\ \ref{fig.fig9}. These value of $F_p$ correspond to critical currents in the absence of the dipole  equal to $0.16J_d,\; 0.32J_d$. For $F_p=0.25\epsilon_0$, the $J_c$ vs. $H$ curve ( Fig.\ \ref{fig.fig9}.a ) and the $J_c$ vs. $\alpha$ curve ( Fig.\ \ref{fig.fig9}.c ) are similar to those for $F_p=0$ (Figs.\ \ref{fig.fig5}.c and \ref{fig.fig5}.a). The differences are that  for $\alpha=180^o$, $J_c$ is weakly dependent on $H$, for $\alpha=146.25^o$ the $J_c$ vs. $H$ curve has a sharp change instead of a discontinuity, and for both  $\alpha=146.25^o,\;135^o$ the $J_c$ vs. $H$ curve is shifted to smaller $H$. These similarities result because as $J$ increases the vortex is pinned at location close to the minimum of  $U^{1v}_{vm}- F_L x$, as shown in Fig.\ \ref{fig.fig9}.b. For $F_p=0.5\epsilon_0$, on the other hand, the strong dependence of $J_c$ on $H$ is destroyed, as shown in Figs.\ \ref{fig.fig9}.c and \ref{fig.fig9}.d. 

In summary, the strong dependence of $J_c$ on $H$ is preserved if the pinning is weak. This means that, for the parameters used in Fig.\ \ref{fig.fig9} the random pinning force must be such that the critical current in the absence of the dipole is  no larger than $\sim 0.2J_d$.

%################################################################################# 
%######################################################################

\section{discussion}
\label{sec.dsc}
The results obtained in this paper are believed to be representative of low-$T_c$ superconducting films with magnetic dipole arrays placed on top. First, the particular  parameters  used, $d\sim z_0\sim \xi$, are typical ones.  For instance, in the experiments with arrays of magnetic dots with permanent magnetization placed on top of superconducting Nb films, reported in Ref.\cite{pann1},  $d=20nm\sim\xi$. The  magnetic dots  are separated from the film by a thin protective layer of thickness $\sim 20nm$, so that the distance from the magnetic dipole to the film is $z_0\sim \xi$.  Second, since the vortex pinning potential  depends on the scaled parameters $m/\phi_0z_0$,  $d/z_0$ and $H\lambda^2/\phi_0$, many superconducting film-dipole array systems are equivalent. 
The London limit used in this paper is valid for vortices in low-$T_c$ films. However, when a magnetic dipole is placed close to the film, it certainly breaks down if the dipole field  destroys superconductivity locally in the film. Roughly speaking, London theory is valid as long as the maximum dipole field at the film is less than the upper critical field,
%############## NEW #########################################################
 that is  $m/z^3_0 < \phi_0/(2\pi \xi^2)$, or $m/(\phi_0z_0) < (z_0/\xi)^2/2\pi$. For the parameters used  above ($z_0=2 \xi$),  this gives $m/(\phi_0z_0) < 0.64$. The values of $m$ used in this paper satisfy this condition. Thus,  the calculations  reported above are within the limits of validity of London theory. 
% ############# END #####################################
The London limit would be a  better approximation if the present calculations were carried out for larger values of $z_0/\xi$. However, the  results for $J_c/J_d$ would be identical to those described above if $m$ and $d$ were scaled by the same factor as  $z_0/\xi$. For instance, if  $z_0\rightarrow 2z_0$, $J_c/J_d$ would remain the same if  $d\rightarrow 2d$ and $m\rightarrow 2m$, but the upper limit of  $m/\phi_0z_0$ for the  validity of the London approximation would increase by a factor of $4$. 
%############## NEW ##################################
The present model also breaks down if $m$ is sufficiently large to create vortices in the film. For the parameters used in the present calculations, the threshold value of $m$ for spontaneous vortex creation is estimated as $m\sim 0.7\phi_0z_0$, using the results of Ref.\cite{gmc1}. This value is larger than $m$ used here. 
% ############# END ######################################################

The simple model discussed here is relevant to arrays of magnetic dots on top of thin superconducting films, provided that: i) the dots are sufficiently far apart to neglect dipole-dipole interactions between them, ii) the number of vortices per dot is small enough, so that each dot pins at most one vortex, and  the vortices are far enough apart to neglect vortex-vortex interactions. The vortices present in the film must be created  by an external field perpendicular to the film surfaces. For the parallel dipole, this field does not affect the dipole orientation. However, this is not the case for the free dipole.  The results obtained in this paper apply to the vortices pinned by the free dipole if the  perpendicular field used to create the vortices is subsequently removed.  

%############################# NEW ##########################################
Now, as an example, the conditions under which the results described above for the parallel dipole apply to a system of experimental interest are examined. The system consists of a typical array of nanomagnets reported in Ref.\cite{ckaw} on top of a thin superconducting film. 
%################# END #######################################
Assuming that $\xi=20nm$, it follows that for  $d=z_0=2\xi$, $\lambda=10\xi$ ( as in Sec.\ \ref{sec.jc}),  $d=z_0=40nm$,   $\lambda=200nm$, and $\phi_0/\lambda^2=500G$. The value $m=0.5\phi_0z_0$ follows if the disk  radius and thickness are chosen respectively as $R\sim 50nm$ and $t\sim 10nm$, and the disk  magnetization is taken as $M\sim 10^2 \mu_B/(nm)^3$. If the  distance between disks in the array is $a\sim 1\mu m$, the dipole-dipole interaction energy , $E_{dd}\sim m^2/a^3$, is small compared with the  vortex pinning potential, since $E_{dd}\sim 10^{-2}[max (U^{1v}_{vm})]$, with $max (U^{1v}_{vm})=0.3\times 4\pi q(m/\phi_0 z_0)$. The values chosen  for the disk radius and thickness, for the magnetization, and for the distance between disks are  typical of those reported in Ref.\cite{ckaw}. For the parameter values described above, the maximum vortex field at the dipole position is $\sim 12 G$. Thus, the results reported in Sec.\ \ref{sec.jc} (Fig.\ \ref{fig.fig5}) predict that for $H<12G\,$, $J_c$ depends strongly on $H$,  like in Fig.\ \ref{fig.fig5}.c, whereas  for $H>12G$, $J_c$ is that for a permanent dipole, and depends only on $\alpha$.

In conclusion then, this paper demonstrates that the interactions between vortices in a thin superconducting film and one freely rotating dipole can be tuned by a magnetic field applied parallel to the film surfaces.  It is shown that the critical current for one vortex pinned by the dipole can be changed  from a few tenths of the depairing current to values larger than the depairing current by changing the applied field.  For  fields much larger than the vortex field, when  the dipole moment is stuck in the field direction, the critical current  changes continously by one order of magnitude when the field is rotated $180^o$, from the direction parallel to the driving force to the direction opposite to it. For fields comparable to the vortex field the critical current is very sensitive to field variations. For a parallel dipole,  very rapid and even discontinuous changes  by as much as one order of magnitude are obtained for the critical current in clean films. It is shown that weak random pinning does not modify substantially this dependence, except for the  discontinuous changes in the critical current which are replaced  by rapid ones. It is suggested that the results  apply to  experiments on magnetic dot arrays on top of  superconducting films. 

\acknowledgments 
Research supported in part by the Brazilian agencies CNPq, CAPES, FAPERJ,  and FUJB.

\end{document}